\documentclass[journal]{IEEEtran}
\usepackage{ifpdf}
\usepackage{cite}
\usepackage{color,soul}
\usepackage[pdftex]{graphicx}
\usepackage[caption=false,font=footnotesize]{subfig}
\graphicspath{{./pic/}}
\DeclareGraphicsExtensions{.pdf,.jpeg,.png}
\usepackage{array}
\usepackage{mathrsfs,amsmath,amsfonts} 
\usepackage{fixltx2e}
\allowdisplaybreaks
\begin{document}
\title{On-line Survival Analysis of Power Electronic Converters Using Step Noise-Cox Processes}
\author{Pourya~Shamsi\IEEEauthorrefmark{2},~\IEEEmembership{Member,~IEEE,}
\thanks{\IEEEauthorrefmark{2} Pourya Shamsi is currently with Missouri University of Science and Technology, Rolla, Missouri 65409 USA (email: shamsip@mst.edu).}}
\maketitle
\thispagestyle{empty}
\pagestyle{empty}
\begin{abstract}
This paper is focused on survival analysis of electrical components. The main goal of this paper is to develop a method for on-line estimation of the Mean Time To Failure (MTTF) of electrical components under dynamic stress levels. The proposed method models the variations of the stress levels as a stochastic process. Hence, a stochastic failure rate function can be developed for each electrical component. Later, this function is used as the underlying rate of a doubly stochastic Poisson process (known as Cox processes). Furthermore, this Cox process is used for on-line estimation of the Mean Residual Life (MRL) using the observed stress levels. The proposed method provides a good estimate of the age and life expectancy of each component. An experimental case study is provided to demonstrate the proposed method.
\end{abstract}


\section{Introduction}
\IEEEPARstart{M}{odern} electrical systems are often operating near their margins of safe operational conditions. In order to reduce the cost and real estate requirements of these systems, traditional safety margins are reduced and these components are exposed to higher stress levels. Hence, dynamic monitoring of the operational state of these components is crucial. Examples of over stressed electrical systems can be found in modern microgrids and hybrid electric vehicles. For instance, in a modern microgrid, distributed power electronic converters interfacing renewable energy sources and storage system are under dynamic variations of the operational conditions. Therefore, thermal and electrical stress level on each component depends on the state of power flow within the microgrid and is not a fixed value. Hence, traditional off-line methods for reliability analysis of these systems cannot provide accurate models for survival analysis of these systems.  

In order to monitor the state of health and age of these components, reliability analysis methods are incorporated. Reliability deals with the ability of each component to perform a required action. However, this qualitative parameter is hard to describe in the mathematical sense. For this reason, this parameter is often described by the life expectancy of the component. For this reason, in this paper, the reliability of a system is analyzed by studying the survival function. Majority of the studies in the field of reliability can be categorized into two branches. \textit{Component level reliability} deals with the probability distribution of the time to failure of a single component \cite{mil,att,ind19}. On the other hand, \textit{system level reliability} studies the reliability of a system whether it is developing a life-time probability distribution model based on the probability distributions of individual components, or studying the state of reliability (operation, safety, health, etc.) of the system using mathematical models \cite{ind18,ind5,ind6,ind2}. The latter is often performed using net structures such as Markov chains and processes. For instance, Markov/hidden Markov models have been widely used to estimate the state (of health) of a system \cite{ind3,ind7}. Another approach is to optimize the system for maximizing the reliability \cite{ind1,ind4,ind10}.

Conventional reliability analysis methods are mainly focused on exponential probability distributions \cite{ind15,ind16,ind13}. This is due to the simplicity in working with the exponential probability distribution. In particular, the Markovian (memoryless) property of this distribution enables for derivation of closed form time domain models for the net (graph) structures developed with this distribution. Markov processes is an example of applications of exponential distribution in a net structure. The base exponential model is calculated based on empirical measurements on accelerated life tests on each component (or component family) \cite{att,mil}. The results from these measurements lead to failure rate functions that can be used for survival analysis \cite{ind14,ind8}. In order to improve the accuracy of the survival model, more complex probability distribution functions can be incorporated \cite{SRTM}. For instance, Weibull distribution is of a great importance in reliability analysis. Distribution of various empirical data from accelerated life tests can be modeled using Weibull distribution function \cite{att}. For this reason, Weibull is often incorporated to increase the accuracy of the reliability model \cite{ind11}.

On the other hand, in conventional methods, after a distribution function is selected (whether it is a simple exponential model or a more real-life-like model), it is tuned based on the expected operational conditions during the life span of the component. In order to be on the safe side, the hazard rates of these models are tuned for the worst case scenarios. However, the electrical components under study will not constantly suffer from the maximum stress levels. Consider a battery storage system in a microgrid. Depending on the power demand within the grid or seasonal conditions, the thermal and electrical stresses over the battery charger will vary. Hence, an on-line method for monitoring the stress levels and aging of the battery charger is of interest. 

This paper presents an on-line method for calculating the age and life expectancy of electrical components. First, a stochastic model for modeling the expected stress levels based on the operational conditions is developed. This model is used as the stochastic acceleration factor in modeling the survival functions of each component. This is performed by introducing Cox processes as the distribution function of the time to failures. The developed process is used for estimating the mean residual life of the component based on each observed stress level. Finally, an experimental study is provided to demonstrate the proposed method. \section{Survival Models}
\noindent In the probability theory, stochastic processes which deal with random collections of points are called point processes. In particular, this paper is incorporating point processes to study random occurrence of incident. In fact, these incidents are failures of electrical components. Renewal processes have been widely used to study the reliability of electrical system. In this paper, a renewal process with a stochastic rate of the process is studied. First, some introductions are required. 
\subsection{Renewal Processes}
\noindent Let $\Omega$ be a sample space and let $\mathcal{F}=\mathfrak{B}(\Omega)$ be the $\sigma$-algebra (Borel $\sigma$-algebra). Let $\mathbb{P}: \mathcal{F}\rightarrow [0,1]$ be a measure on $\mathcal{F}$. The quartet $(\Omega,\mathcal{F},\{ \mathcal{F}_t \}_{t \ge 0}, \mathbb{P})$ is a filtered probability space with a natural filtration of $\{ \mathcal{F}_t \}_{t \in I}$, where $I \in \mathbb{R}^{+}$ (practically, $I \in \mathbb{N}$) is a totally ordered index set. Although the filtration might not be of great importance for the analysis of a Homogeneous Poisson Process (HPP), it is required for the later sections. The history is defined as
\begin{flalign}
&{\mathcal{F}_\infty } := { \vee _{\tau  \ge 0}}{\mathcal{F}_\tau }&\\
&{\mathcal{F}_\tau } = \{ A \in \mathcal{F} | A \cap \{ \tau  \le t\}  \in {\mathcal{F}_t },\forall t \ge 0 \} &
\end{flalign}
On the other hand, let $\mathbf{X}$ be a topological space with its Borel $\sigma$-algebra $\mathfrak{B}(\mathbf{X})$ to define random variables $x \in \mathbf{X}$ as
\begin{flalign}
&x:(\Omega, \mathcal{F})\rightarrow(\mathbf{X},\mathfrak{B}(\mathbf{X}))&
\end{flalign}

Consider an orderly counting process $\mathcal{N}(t)_{t > 0}$ which is adapted to the filtration $\mathcal{F}_t$. In this process, $\mathcal{N}(t)$ represents the number of incidents during the time interval $(0,t]$ while $\mathcal{N}(0)=0$. For now, it is assumed that the process has independent deterministic increments. Hence, the infinitesimal probabilities of the process are defined as
\begin{flalign}
\label{_eqinf}
&\Pr \{ \mathcal{N}(t + \delta ) -\mathcal{N}(t) = 1|\mathcal{F}_t\} = w(t)\delta  + o(\delta )&\\
&\Pr \{ \mathcal{N}(t + \delta ) -\mathcal{N}(t)> 1|\mathcal{F}_t\} = o(\delta )&
\end{flalign}
where $w(t)$ is the renewal density (in probability theory) or the rate of occurrence of failure (ROCOF) (in reliability theory). This function defines the rate of the process. Before studying the process any further, let $T:\Omega \rightarrow \mathbb{R}^{+}$ denote a random variable representing the interval of the time till the first incident in the process (i.e. $(0,T_1]$). The survival function of this incident is defined as
\begin{flalign}
&S_{T_1}(t)=1-F_{T_1}(t)=\Pr \{T_1>t\}&
\end{flalign}
This function represents the probability of the incident throughout the time. Hence, the probability of having a failure at time $(t,t+\delta t]$ while the unit has survived till $t$ is derived using
\begin{flalign}
\label{_zt}
&\lambda (t) = \Pr {\{ t < T_1 \le t + \delta |t < T_1 \}/\delta  |_{\delta  \to 0}} = \tfrac{{f_{T_1}(t)}}{{S_{T_1}(t)}}&
\end{flalign}
where $f_{T_1}(t)=\dot F_{T_1}(t)$ is the probability density function. $\lambda(t)$ is defined as the failure rate function or the Force Of Mortality (FOM). One should avoid interchanging the ROCOF and the failure rate function since the failure rate function represents the hazard rate of a single incident whereas ROCOF defines the rate of incidents in a process. Further discussion is available in \cite{RSRM}.
\subsection{Improvements to the Failure Rate Function}
Conventionally, the majority of studies on reliability assessment in electrical systems consider a constant failure rate function for each component. This failure rate function is known as the {\textit{failure rate}} for short. Using a fixed $\lambda(t)=\lambda$, the survival function is derived using 
\begin{flalign}
&S(t) = {e^{ - \int_0^t {\lambda dt} }} = {e^{ - \lambda t}}&
\end{flalign}
\begin{figure}[tb]
\centering
\includegraphics[trim = 0.2in 0.2in 0.2in 0.2in, clip, width=2.8in]{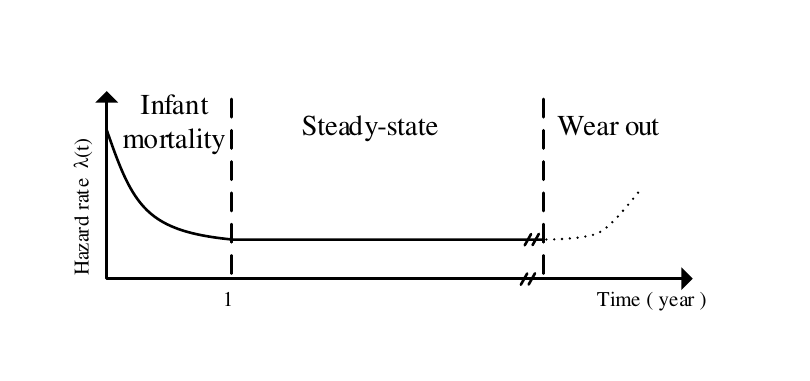}
\caption{The bathtub failure rate function.}
\label{_bathtub}
\end{figure}
Memory-less property of exponential distributions results in simplicity of analytical studies. However, it has been observed that a constant failure rate function cannot represent the measured hazard rates in a practical application. Many studies suggest using a bathtub curve to include the infant mortality and wear-out of individual components. For instance, Fig. \ref{_bathtub} illustrates a recommended hazard rate for modeling electrical components by the American telecommunication provider, AT\&T \cite{att}. This curve does not consider the wear-out period. However, the infant mortality is included using Weibull distributions. This hazard rate models the infant mortality for nearly a year (10 thousand hours). Afterwards, it is reduced into a constant failure rate model (or a Weibull with $\alpha \simeq 0^+$ for more accurate results). The survival function of such failure rate model is calculated as
\begin{flalign}
\label{_wei}
&S(t) = \left\{ {\begin{array}{*{20}{l}}
{{e^{\tfrac{{ - \lambda_b T_0^\alpha {t^{1 - \alpha }}}}{{1 - \alpha }}}}}&{t < {T_0}}\\
{{e^{\tfrac{{ - \lambda_b T_0}}{{1 - \alpha }} - \lambda_b (t - {T_0})}}}&{{T_0} \le t}
\end{array}} \right.&
\end{flalign}
where $T_0 \in \mathbb{R}^{+}$ is the base transition time ($10^4$ hours $\sim$ 1 year). Several studies have developed tables for components commonly used in electrical systems \cite{att,mil}. $\alpha \in [0,1)$ is the empirically measured shape parameter of the Weibull distribution. The data regarding the shape parameters are available based on accelerated life test studies. One can observe that (\ref{_wei}) is not a standard $(\alpha_w,\lambda_w)$ Weibull distribution. However, (\ref{_wei}) is generated with the change of variables of $\alpha_w=1-\alpha$ and $\lambda_w=(\lambda_b T_0^{\alpha} / (1-\alpha))^{1/(1-\alpha)}$.

The base failure rate, $\lambda_b \in \mathbb{R}^{+}$, in (\ref{_wei}) is calculated based on results from empirical measurements. It should be noted that the results from different measurements do not necessarily lead to same failure rates. This is due to the components under test as well as type of the accelerated testing. Let the component under study be an npn Bipolar Junction Transistor (BJT). In order to simplify the study, a 45V 0.5A switch with a transition frequency (i.e. $f_T$) of 100 MHz is selected. It should be noted that the presented method is similar for every electrical components and the selected transistor is to provide a case study for clarifying the method. For Weibull distribution, the operational failure rate function of this switch can be estimated using
\begin{flalign}
&\lambda= A_a(\alpha) T_0^{\alpha}{\lambda _b}&
\end{flalign}
where $\lambda_b$ is the base failure rate with a value of $20 \times 10^{-9}$ failures per hour for the selected BJT. $\alpha$ is $0.6$ for the selected switch. $A_a(\alpha) \in [1,\infty)$ is the stress related aging factor (acceleration factor) which is calculated using
\begin{flalign}
\label{_eqacc}
&{A_a}(\alpha) = ({A_T}{A_e}{A_{en}})^{(1-\alpha)}&
\end{flalign}
where $A_{en}=1$ is the environmental factor. $A_e$ is the electrical factor and depends on the type of the component. For the case of a BJT, this parameter can be calculated using
\begin{flalign}
&{A_e} = \max \{ 1,0.47{e^{3\tfrac{{{v_{CE}}}}{{v_{CE}^*}}}}\}&
\end{flalign}
where the breakdown voltage of the switch is $v_{CE}^* = 45$ while the operational voltage is $v_{CE}$. Furthermore, the thermal coefficient $A_T$ can be derived using
\begin{flalign}
&{A_T} = {e^{\tfrac{E_a}{K_b}(\tfrac{1}{T_r} - \tfrac{1}{T_o})}}&
\end{flalign}
where $E_a$ is the activation energy of 0.4 eV. $K_b$ is the Boltzmann constant. $T_r$ is the reference temperature of 313 K while $T_o$ is the ambient operational temperature. Hence, the failure rate function for this case study can be approximated by linearization near the nominal operation point as
\begin{flalign}
\label{_lin}
&\lambda  \simeq&\\
& \left\{ {\begin{array}{*{10}{l}}
{{\lambda _b}T_0^\alpha {{8.4}^{1 - \alpha }}(1 + (1 - \alpha )(0.06\tilde v + 0.04{{\tilde T}_a}))t^{-\alpha}}&{t < {T}}\\
{{\lambda _b}{{8.4}}(1 + (0.06\tilde v + 0.04{{\tilde T}_a}))}&{{T} \le t}
\end{array}} \right.&\nonumber
\end{flalign}
near the operation condition of $\bar v=30\;\rm{V}$ and $\bar {T_o}=333\;\rm{K}$. It should be noted that under accelerated conditions, Weibull distribution will transition into an exponential distribution at time $T<T_0$ (this transition occurs sooner than the time $T_0$ defined by failure rate tables such as AT\&T handbook). Therefore, $T$ can be calculated by finding the point where Weibull and exponential failure functions meet. In a conventional survival analysis, the failure rate function is calculated based on the nominal operation conditions. However, in a practical application, the aging factors (i.e. $\tilde v$ and $\tilde {T_o}$) vary with time. For instance, seasonal stochastic variations of temperature can lead to accelerated aging of a component. Hence, to achieve a more accurate failure rate function, on-line calculation of the acceleration factor is of interest. Moreover, this suggests that the acceleration factor has a stochastic nature. One solution is to incorporate doubly stochastic Poisson processes known as Cox processes. Cox processes were first introduced by D. R. Cox in observation of patterns in empirical data \cite{cox0,dspp}. This process benefits from random failure rate functions. Hence, the \textit{rate} of this point process itself is a stochastic process. Cox processes have been widely used in economic studies \cite{cox1,cox2,cox3,cox4,cox5,cox6}.
\subsection{Step Noise-Cox Process}
\noindent It was mentioned that the aging factors governing the gain of a failure rate function have stochastic variations. The failure rate function of a BJT was developed in (\ref{_lin}). It can be observed that both inputs of temperature and voltage have a same effect on the overall failure rate function. If variations of multiple parameters is of interest, due to linear nature of (\ref{_lin}), the random variables are summed and hence, individual characteristic functions can be multiplied. For this reason, this section studies the stochastic variations of one parameter (voltage). Variations of the voltage of the switch can be modeled using a simpler stochastic process.
\begin{figure}[tb]
\centering
\includegraphics[trim = 0in 0in 0in 0in, clip, width=2.5in]{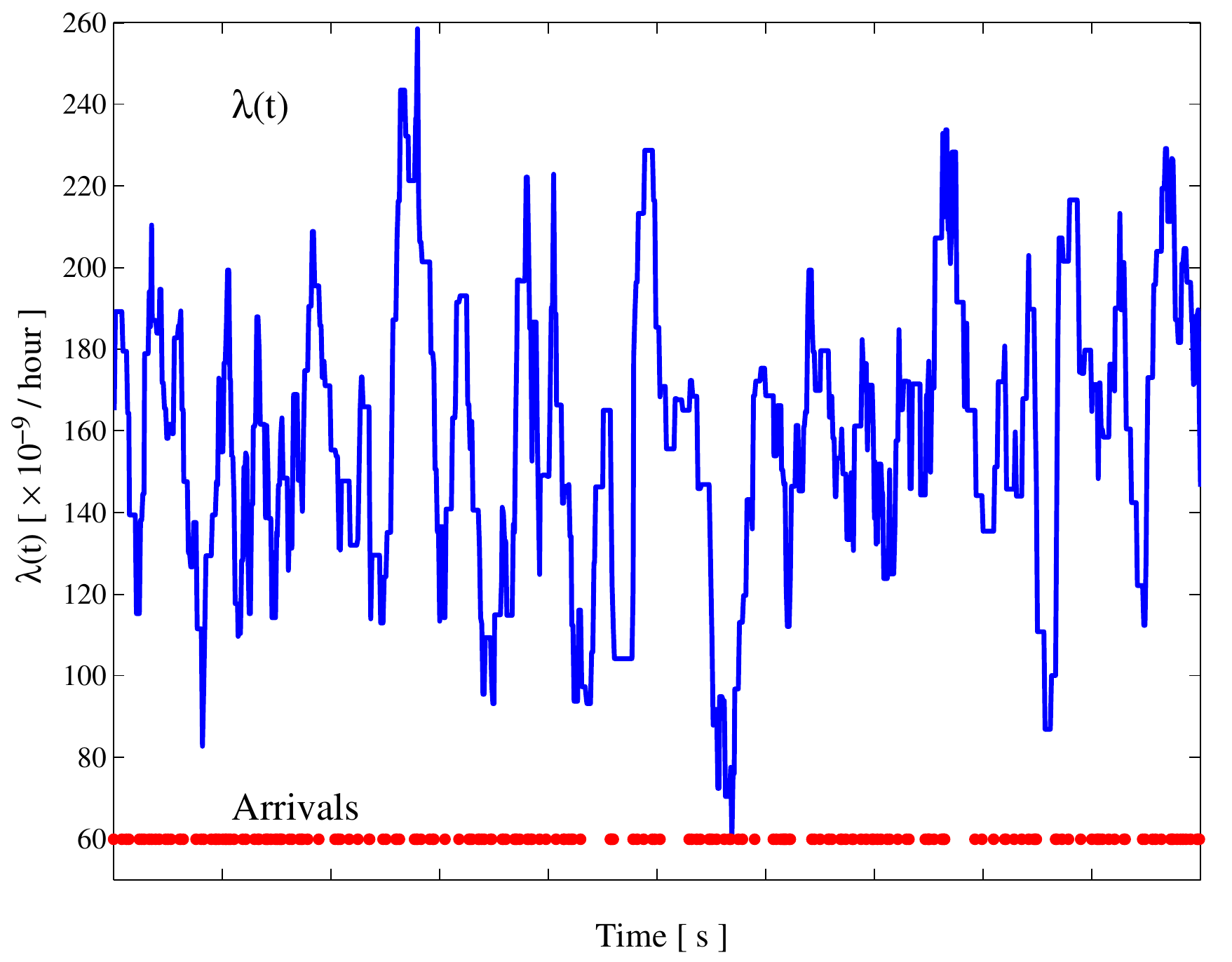}
\caption{A sample path of the failure rate function $\lambda_s(t)$.} 
\label{_vt}
\end{figure}
Let
\begin{flalign}
\label{_eqvs}
&{\tilde v}_{CE} = {v_t} = {\nu _0} + \sum\limits_{k = 1}^{{\mathcal M}_t} { \nu _k (u(t-\mathsf{S}_k)-u(t - \mathsf{S}_k - \tau ))}&
\end{flalign}
be the stochastic process estimating the variations of the operational voltage. In this process, random variable $\nu _k : \Omega_{\nu} \rightarrow \mathbb{R}$ represents a random voltage step on the voltage of the switch while $\nu_0=0$. The above process is similar to a compound Poisson. This paper refers to Poisson processes with the above rate function as {\bf{step-noise cox}} processes. Here, $\{\mathcal {M}_t\}_{t>0}$ is a homogeneous Poisson process with intensity $\rho $. $\tau \in \mathbb{R}^+$ is the length of each step noise in time. $\mathsf{S}_k$ is the time until the $k$-th arrival and is calculated using
\begin{flalign}
&{\mathsf{S}_k} = \sum\limits_{k = 1}^{{{\mathcal M}_t}} {{\mathsf{T}_k}} &
\end{flalign}
where $\mathsf{T}_k:\Omega \rightarrow \mathbb{R}^{+}$ is the $k$-th inter-occurrence time. The stochastic process defined in (\ref{_eqvs}) models the random behavior of the voltage of the switch, therefore, the stochastic variations of the accelerated failure rate function can be modeled using (\ref{_lin}) and (\ref{_eqvs}) combined as
\begin{flalign}
\label{_eqlam}
&\lambda  \approx {\lambda _0}{t^{ - \alpha }} + \kappa \sum\limits_{k = 1}^{{{\cal M}_t}} {{\nu _k}(1 - u(t - {\mathsf{S}_k} - \tau ))} {t^{ - \alpha }}&
\end{flalign}
It should be noted that $\lambda_0$ is the starting value of the $\lambda$ while $\kappa$ represents the gain of the perturbations with respect to (\ref{_lin}). Also, in the previous example, $\lambda_0=165 \times 10^{-9}$ and $\kappa = 165 \times 0.066 \times 10^{-9}$. This function can be derived for the case of exponential distribution by setting $\alpha=0$. Additionally, due to the characteristics of the compound Poisson process in (\ref{_eqlam}), replacing $u(t - {\mathsf{S}_k})$ with $1$ does not change the function. 

A sample path for the stochastic behavior of the exponential failure rate function is illustrated in Fig. \ref{_vt}. In this figure, the random steps of the voltage, $\nu _k$, have a uniform distribution on $[-5,5]$. It can be deduced from this simple example that variations of the voltage of the selected BJT can change the average number of failures from 60 to 280 incidents per million hours of operation. The reason for proposing a step-noise process in this paper is the behavior of power electronic converters. Consider a source with a series impedance. Due to the step changes in the duty cycle of the converters, the voltage drop over the series impedance of each source is in the form of a step. These voltage steps are highly observable in utilization of a stair-case method for MPP tracking of photovoltaic solar panels. 

Due to the stochastic variations of the failure rate function, (\ref{_eqlam}) is denoted as $\lambda(t)$. The process is assumed to be operating on the same probability space of $(\Omega,\mathcal{F},\{ \mathcal{F}_t \}_{t \ge 0}, \mathbb{P})$ which was defined earlier and is measurable. The process $\lambda(t)>0$ is adapted to $\mathcal{F}_t^{\lambda}$ (and in fact is predictable). The counting process generated by this intensity is called a doubly stochastic Poisson process or Cox process. The necessary condition for validity of the following studies is to eliminate any explosions in the process. Hence, let
\begin{flalign}
& \int_0^t {{\lambda (s) }ds < \infty,\; \lambda(t)>0} \;,t \ge 0&
\end{flalign}
And for $0 \le s \le t$ and $\forall u  \in \mathbb{R}$ while having the full realization of $\lambda (t)$ as $\mathcal{F}_{\infty}^{\lambda}$, the conditional characteristic function of the resulting Poisson process is defined as \cite{cox2}
\begin{flalign}
&\mathsf{E}[{e^{- \jmath u ({\mathcal{N}_t} - \mathcal{N}_s)}}|{\mathcal{F}_s}\vee {\mathcal{F}_{\infty}^{\lambda}}] = {e^{({e^{-\jmath u }} - 1)\int_s^t {{\lambda (u)}du} }}&
\end{flalign}
which is quite well-known for Poisson processes and therefore, the probability of having $k$ failures can be calculated using
\begin{flalign}
\label{_eqprob}
&\Pr \{ {{\mathcal N}_t} - {{\mathcal N}_s} = k|{\mathcal{F}_{\infty}^{\lambda}}\}  = \tfrac{{{{\left( {\int_s^t {{\lambda (u)}du} } \right)}^k}{e^{ - \int_s^t {{\lambda (u)}du} }}}}{{k!}}&
\end{flalign}
where ${\mathcal{F}_{\infty}^{\lambda}}$ denotes the realization of the stochastic process generating the failure rate function. In order to benefit from this process in a practical application, the dependency to the future has to be eliminated (i.e. ${\mathcal{F}_{\infty}^{\lambda}}$). Also note that
\begin{flalign}
\Lambda(t)& = \int_0^t {{\lambda (u)}du}&\\
&=\int_0^t t^{-\alpha}({\lambda_0+\kappa  \sum\limits_{k = 1}^{{\mathcal M}_t} { \nu _k (1-u(t - \mathsf{S}_k - \tau ))}})dt&\nonumber
\end{flalign}
is the aggregated process, $\Lambda_t $, which affects the characteristic function of the Cox process. Additionally, the processes in this paper are optional c\`{a}dl\`{a}g processes. $\Lambda_t$ can be integrated directly as a new compound Poisson process 
\begin{flalign}
&\Lambda(t) = \lambda_0 t^{1-\alpha}/(1-\alpha)+\kappa  \sum\limits_{k = 1}^{{\mathcal M}_t} { \nu _k \tau \mathsf{S}_k^{-\alpha}}=\Lambda_1+\Lambda_2&
\end{flalign}
where $(t^{1-\alpha}u(t-\mathsf{S}_k)-(t-\tau)^{1-\alpha}u(t-\mathsf{S}_k-\tau))/(1-\alpha)$ is approximated as $\tau \mathsf{S}_k^{-\alpha}$ since $\tau \ll t$. In order to derive the characteristic function of the Cox process, extension of Fokker-Planck-Kolmogorov for jump-diffusion based on It\^{o} calculus is incorporated. This extension which is useful for jump processes was originally developed for piecewise deterministic Markov processes \cite{generator1,generator2,PPMP}. Using this concept, the infinitesimal generator of the process $\Lambda_2$ can be calculated for the characteristic function ($\Lambda_1$ is derived in the exact same way, however, since $\Lambda_2$ is more complex, we demonstrate derivation of this part and we leave the first part to the reader) 
\begin{flalign}
&f(\Lambda_2 ,t) = \mathsf{E}[e^{- \jmath u \Lambda _2}|{\Lambda _{2_0}}]&
\end{flalign}
using
\begin{flalign}
\label{_eqa_}
Af(\Lambda_2& ,t)= \tfrac{{\partial f}}{{\partial t}} +\tfrac{{\partial f}}{{\partial \Lambda_2}} \tfrac{{\partial \Lambda_2}}{{\partial t}}& \nonumber\\
&+\rho[\int_{\mathcal{D}(G)} {f(\Lambda_2  + \kappa \tau \zeta ,t)dG(\zeta )}  - f(\Lambda_2 ,t)]&
\end{flalign}
since $\partial \Lambda_2 /\partial t=-\alpha t^{-1} \Lambda_2$, with a change of variable of $z=\Lambda_2 t^\alpha$, the partial derivative of $f(z,t)$ with respect to $z$ has the terms $\tfrac{{\partial f}}{{\partial z}} (\tfrac{{\partial z}}{{\partial \Lambda_2}}\tfrac{{\partial \Lambda_2}}{{\partial t}}  +\tfrac{{\partial z}}{{\partial t}} )=\tfrac{{\partial f}}{{\partial z}} (-t^\alpha \times \alpha \Lambda_2 t^{-1} +\alpha \Lambda_2 t^{\alpha-1}) =0$ and therefore, this change of variable is very useful by transforming the infinitesimal generator into
\begin{flalign}
Af(z ,t)&= \tfrac{{\partial f}}{{\partial t}} +\rho [\int {f(z +  \kappa \tau \zeta t^\alpha ,.)dG(\zeta )}  - f(.,.)]&
\end{flalign}
and hence, $df$ is calculated as
\begin{flalign}
\label{_eqa}
df(z ,t) &= Af(z,t) & \\
&+ [\int {f(z +  \kappa \tau \zeta t^\alpha,t)dG(\zeta )}  - f(z ,t)]d{\tilde N_t}&\nonumber
\end{flalign}
where $d{\tilde N_t}$ is the compensated jump process (i.e. $dN-\lambda dt$). Therefore, based on Watanabe’s characterization of a Poisson process, this term is a martingale. $G(x)$ is the probability kernel and in general is a function of time as well (i.e. $G(x,t)$). In the special case studied in this paper, $G(.)$ represents a probability distribution of random voltage steps $\nu$. 

It is very important to note that the infinitesimal semi-group is being used for a non-Markovian process. Therefore, care should be taken to include the calendar time in the derived equation and not the time since the last observation $t-s$. Therefore, by $t$, all of the above and following equations refer to the calendar time $t$ starting from $t_0=0$ which is the first time the device was installed. In this application, the transition semi-group $T_{t,s}$ should be used and not the Markovian transition semi-group $T_t$. Moreover, $A_t$ is written as $A$ with an abuse of notation.

With respect to (\ref{_eqa}), a reference function in the form of
\begin{flalign}
&f(\Lambda_2 ,t) =  H{e^{\int_0^t {\rho \left( { \int {{e^{ - \jmath u \zeta \kappa \tau k^{-\alpha}}}dG(\zeta )} -1} \right)dk} }}&
\end{flalign}
is selected. Moreover, it should satisfy the boundary condition and should be multiplied with the characteristic function of $\Lambda_2$. Hence,
\begin{flalign}
\label{_prcox1}
\mathsf{E}[e^{- \jmath u \Lambda _t}|{\lambda _0}]&= {e^{ - \jmath u t^{1-\alpha} \lambda_0 /(1-\alpha)}}&\\
&\times {e^{\int_0^t {\rho \left( { \int {{e^{ - \jmath u\zeta \kappa \tau k^{-\alpha}}}dG(\zeta )}-1 } \right)dk} }}&\nonumber
\end{flalign}
Now, we can observe the characteristic function of the Cox process. First, it was mentioned that due to the conditional on the future, the characteristic function of the Cox process cannot be directly calculated. Hence, the expected characteristic function can be derived using
\begin{flalign}
\label{_prcox2}
&\mathsf{E}[{e^{- \jmath u ({\mathcal{N}_t} - \mathcal{N}_s)}}|{\mathcal{F}_s}\vee {\mathcal{F}_{s}^{\lambda}}] = \mathsf{E}[{e^{({e^{-\jmath u }} - 1)\int_s^t {{\lambda (u)}du} }}]&
\end{flalign}
which can be combined with (\ref{_prcox1}) (using a change of variable of $-\jmath u = (e^{-\jmath u}-1)$ and assuming $s=0$) to derive
\begin{flalign}
\label{_cox1}
\mathsf{E}[e^{- \jmath u {\mathcal{N}_t}}|{\lambda _0}]&= {e^{({e^{-\jmath u }} - 1)t^{1-\alpha} \lambda_0/(1-\alpha)}}&\\
&\times {e^{\int_0^t {\rho \left( { \int {{e^{ ({e^{-\jmath u }} - 1)\zeta \kappa \tau k^{-\alpha}}}dG(\zeta )}-1 } \right)dk} }}&\nonumber
\end{flalign}
for ${\mathcal{N}_0}=0$ and the derivation of the characteristic function of the proposed step noise-Cox process is finished. 

Following relations are observed based on (\ref{_prcox1})
\begin{flalign}
\label{_cox2}
&\mathsf{E}[{\lambda_t}|{\lambda _0}]=(\lambda_0+\rho \tau \kappa_0 \mathsf{E}_G[\nu])t^{-\alpha}&\\
\label{_cox2_}
&\mathsf{E}[{\Lambda_t}|{\lambda _0}]= (\lambda_0  +  \rho \tau \kappa_0 \mathsf{E}_G[\nu] )t^{1-\alpha}/(1-\alpha)&
\end{flalign}
It is very important to note that the compound Poisson process in $\lambda$ is windowed. Otherwise, the expected value of the process would be $\mathsf{E}[{\lambda_t}|{\lambda _0}]=\lambda_0 t^{-\alpha} + \rho \kappa_0 \mathsf{E}_G[\nu] t^{1-\alpha}$ and $\mathsf{E}[{\Lambda_t}|{\lambda _0}]=\lambda_0 t^{1-\alpha}(1-\alpha) + \rho \kappa_0 \mathsf{E}_G[\nu] t^{2-\alpha}/(2-\alpha)$. However, in this paper, each new step noise will survive only for a duration of $\tau$. Based on (\ref{_cox1}), ROCOF $w(t)=\mathsf{E}[\mathcal{N}_t|t_0=0,\; \lambda_0]$ is calculated by getting the derivative of the characteristic function at $u\rightarrow 0$. Therefore, 
\begin{flalign}
\mathsf{E}[\mathcal{N}_t|t_0,\lambda_0]&=\lambda_0 \dfrac{t^{1-\alpha}}{(1-\alpha)}+\rho \kappa\tau \iint \zeta k^{-\alpha} dG(\zeta) dk&
\end{flalign}
and $\Pr \{ {{\mathcal N}_t}= 0 \}=S(t)=\Pr \{ T_f>t \}$, hence
\begin{flalign}
\label{_eqn0}
\Pr \{ T>t \} &= e^{ - t^{1-\alpha} \lambda_0 / (1-\alpha)}  e^{-\rho \kappa \tau t^{1-\alpha}/(1-\alpha) \mathsf{E}_G[\nu] }&
\end{flalign}
for ${{\mathcal N}_0}= 0$. By suppressing $G$ (i.e. letting $f_G(\nu)=\delta(\nu - 0)$), the model simplifies to Weibull as
\begin{flalign}
\Pr \{ T>t \} &= {e^{ - {\lambda _0} t ^{1-\alpha}/(1-\alpha)}}&
\end{flalign}
and by letting $\alpha\searrow0$, exponential distribution for modeling reliability of electrical components is derived. These two models are traditionally used for reliability analysis of electrical components. Hence, the proposed model is more detailed and can fully contain the conventional models.

In this section, the stochastic failure model of an electrical component based on a step noise-Cox process was developed. Next section will use this model for on-line survival analysis of a component.
\section{On-line Survival Analysis}
\noindent In conventional methods, the failure rate functions for a component are calculated using the expected information from the operational conditions. However, in a real life application, these expectations do not necessarily occur. For example, consider a power converter deployed to a location. This location has an average summer temperature of 100$^{\circ}$ F while the average winter temperature is 30$^{\circ}$ F. It can be observed that the change in these operational conditions are not included in the off-line failure rate function calculations. For this reason, this paper is proposing {\textit{On-line}} or {\textit{Real-time}} survival analysis by updating the failure rate functions timely with respect to the observed operational conditions. For this purpose, the concept of residual life is used
\begin{flalign}
&{\rm{Rl}} \{ x|s\}  = \Pr\{ T > x + s|T > s\} &
\end{flalign}
where the inter-occurrence time, $T$, is the time to the first failure (more accurately, $T$ is the time to a failure with respect to the preceding failure). $s>0$ is the time that the process was observed. Therefore, $s$ is the moments that the failure rate function is recalculated based on the new operational conditions. $x>0$ is the time in the future. Therefore, if we define a global time (calendar time) $t$, then at any moment after $s$, $t=s+x$. Moreover, it was assumed that the variations of the failure rate function has a stochastic nature. On the other hand, at an observation time, $s$, the failure rate function prior to $s$ has been {\textit{observed}}. Hence, the measured hazard rate values have a deterministic nature with respect to filtration $\mathcal{F}_s$. Therefore, the residual life of the process is 
\begin{flalign}
\label{_eqnrl}
{\rm{Rl}} \{ x|s\}  & = {e^{ - ((x+s)^{1-\alpha}-s^{1-\alpha})\lambda_s/(1-\alpha)}}&\\
&\times e^{-\rho \kappa_s \tau ((x+s)^{1-\alpha}\mathsf{E}_G[\nu] - s^{1-\alpha} \mathsf{M})/(1-\alpha)}&\nonumber
\end{flalign}
where $\mathsf{M}$ is the mean value of the observed fluctuations on the failure rate function. In many applications, it is assumed that $\mathsf{M}$ is zero. However, claim is that in many applications, this value is not zero since the assumptions on stress factors do not necessarily occur and stresses can be higher or lower than the assumed values. In order to expand this definition for the bathtub curve of Fig. \ref{_bathtub}, the residual life can be calculated using 
\begin{flalign}
\label{_eqres}
{\rm{Rl}} \{ x|s\}  &= {1_{x + s \le T_t}}{e^{ - ((x+s)^{1-\alpha}-s^{1-\alpha})\lambda_s/(1-\alpha)}}&\\
&\times e^{-\rho \kappa_s \tau ((x+s)^{1-\alpha}\mathsf{E}_G[\nu] - s^{1-\alpha} \mathsf{M})/(1-\alpha)}&\nonumber\\
& + {1_{x + s > T_t,s < T_t}}{e^{ - ((T_t)^{1-\alpha}-s^{1-\alpha})\lambda_s/(1-\alpha)}}&\nonumber\\
&\times e^{-\rho \kappa_s \tau ((T_t)^{1-\alpha}\mathsf{E}_G[\nu] - s^{1-\alpha} \mathsf{M})/(1-\alpha)}&\nonumber\\
&\times {e^{ - (x+s-T_t)\lambda_s^0}}\times e^{-\rho \kappa_s^0 \tau (x+s-T_t)\mathsf{E}_G[\nu]}&\nonumber\\
& + {1_{s \ge T_t}}{e^{ - x{\lambda_s^0}}}\times e^{-\rho \kappa_s^0 \tau ((x+s)\mathsf{E}_G[\nu] - s\mathsf{M})}&\nonumber
\end{flalign}
where $\lambda_s^0$ is the modified failure rate function at the moment of transition, $\lambda_s$ is the modified failure rate function based on the operational conditions, $\kappa_s$ and $\kappa_s^0$ are the modified gains of the random small signal variations of the voltage and temperature, respectively. These parameters for the case study of the BJT can be calculated using
\begin{flalign}
\label{_eqacl}
&{{\bar A}_a}(n) = 0.47{e^{\tfrac{{3v_{CE}^s}}{{v_{CE}^*}} + \tfrac{{0.4eV}}{{{K_b}}}(\tfrac{1}{{313}} - \tfrac{1}{{T_a^s}})}}&\\
&{\lambda _s} = {\lambda _b}{{\bar A}_a}^{1 - \alpha }(n)T_0^\alpha &\\
&{\kappa _s}_v = {\lambda _s}(1 - \alpha )(\tfrac{{3 }}{{v_{CE}^*}})&\\
&{\kappa _s}_T = {\lambda _s}(1 - \alpha )( \tfrac{{0.4eV}}{{{K_b}{{(T_a^s)}^2}}})&\\
&\lambda _s^0 = {\lambda _b}{{\bar A}_a}(n)&\\
&\kappa_{s_v}^0 = \lambda _s^0(\tfrac{{3}}{{v_{CE}^*}})&\\
&\kappa_{s_T}^0 = \lambda _s^0(\tfrac{{0.4eV}}{{{K_b}{{(T_a^s)}^2}}})&
\end{flalign}
where $\bar A_a$ is the base acceleration factor calculated at the observation time $s$ (based on the observed operational conditions). Indexes $v$ and $T$ determine the $\kappa$ for variations of voltages and temperatures, respectively. $T_t$ is the transition time. This value represents the remaining time to pass the infant mortality band. For the $n$-th observation period, this value can be calculated using
\begin{flalign}
&T_t = s + \tfrac{1}{{\bar A_a}(n)}({T_0} - T_{Age})&\\
&T_{Age}=T_{smp}\sum\limits_{k = 1}^{n - 1} {\bar A_a(k)}&
\end{flalign}
where $T_{smp}$ is the time period between each on-line residual life calculation. $T_0$ is the base infant mortality period which was described in the previous section (i.e. $10^4$ hours). $T_{Age}$ represents the age of the component which is affected by the acceleration factors. Therefore, a component can be older than the elapsed calendar time. $A_a(k)$ is the aging acceleration factor estimated at the $k$-th on-line estimation period and was described in (\ref{_eqacc}). $A_a(n)$ is calculated using (\ref{_eqacc}) based on the average voltage and temperature data measured since the last estimation period. 

\begin{figure}[tb]
\centering
\includegraphics[trim = 0in 0in 0in 0in, clip, width=2.5in]{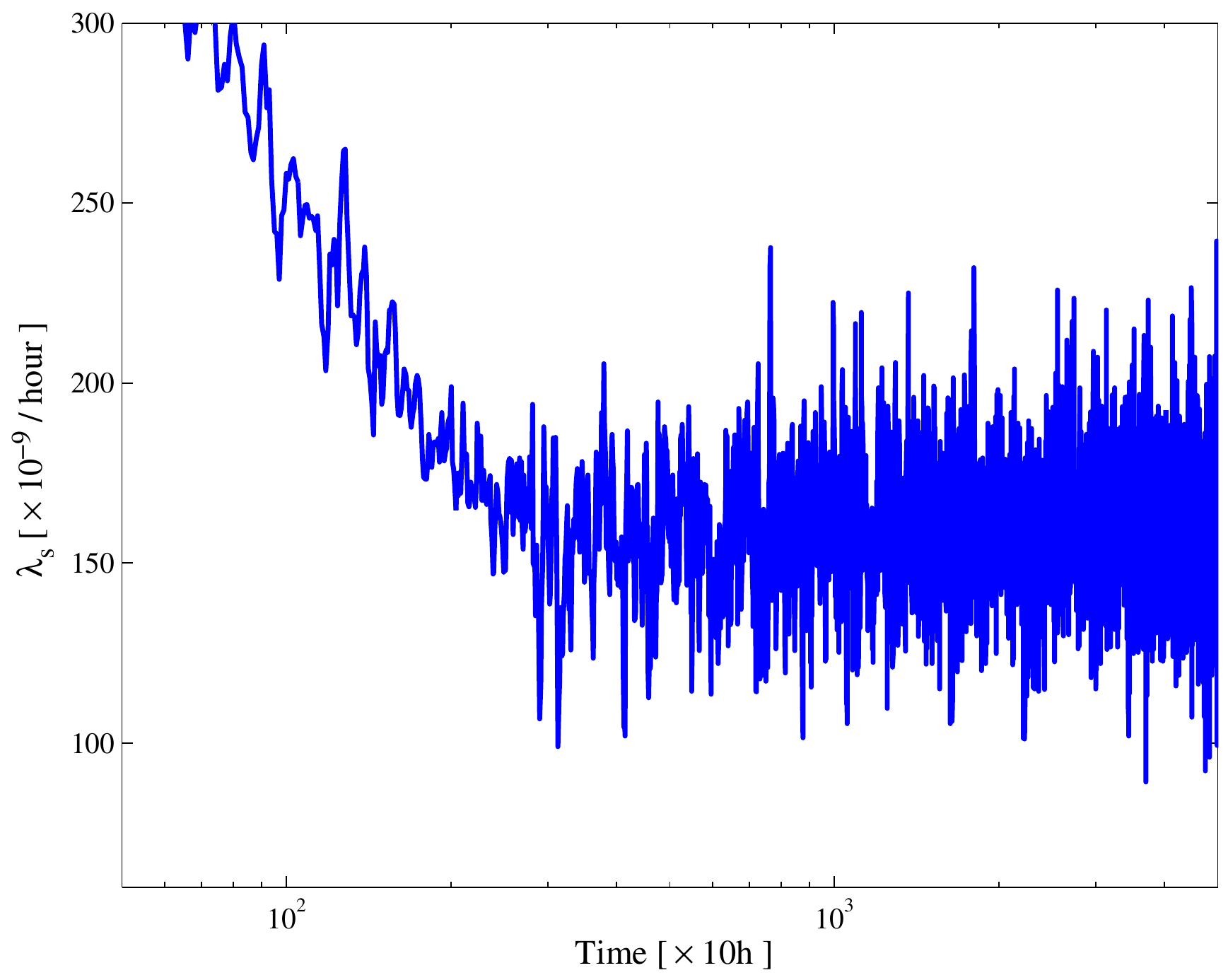}
\caption{A 5 year sample path of the failure rate function $\lambda_s$.} 
\label{_vt1}
\end{figure}

In order to illustrate the transition between the Weibull and exponential (infant mortality and constant failure rate) regions, a sample path for the failure rate function is illustrated in Fig. \ref{_vt1}. In this figure, the average acceleration factor is kept constant at 4. Therefore, the base transition time is reduced from 1 year ($10^4$ hours) to 2500 hours. In addition, it can be observed that the average failure rate is constant and stochastic fluctuations with zero mean are affecting the failure rate function. Moreover, impacts of the acceleration factors are lower during the infant mortality band.

An important note regarding (\ref{_eqacl}) is that in this equation, the variations of the voltage and temperature are included at the same time. However, the stochastic behavior of the failure rate function was derived considering the variations of the voltage and not the temperature. If the variations of all parameters occur at a same arrival time (which does not affect the accuracy of the results of this paper), then all of the derived equations remain the same. However, characteristic functions of the process related to temperature and the process related to voltage should be multiplied. This results in a survival function in the form of 
\begin{flalign}
\Pr \{ T>t \} &= e^{ - t^{1-\alpha} \lambda_0 / (1-\alpha)} &\\
&\times e^{-(\rho_\nu \kappa_\nu \tau_\nu  \mathsf{E}_{G_\nu}[\nu]+\rho_T \kappa_T \tau_T  \mathsf{E}_{G_T}[T])t^{1-\alpha}/(1-\alpha) }&\nonumber
\end{flalign}

Using (\ref{_eqres}), the Mean Residual Life (MRL) of a component can be calculated by
\begin{flalign}
&{\rm{MRL}}(s)= \mu (s) = \int_0^\infty  {{\rm{Rl}} \{ x|s\} dx}&
\end{flalign}
In general, a closed form solution to this integral is not available. However, numerical approximation of this integral can be calculated using Riemann sums at each observation moment. MRL of the (\ref{_eqres}) can be obtained using
\begin{flalign}
\mu (s) &= {1_{s \le {T_t}}} [\int_0^{{T_t} - s} {\{ {e^{ - \left( {{{(x + s)}^{1 - \alpha }} - {{s}^{1 - \alpha }}} \right){\lambda _s}/(1 - \alpha ) }}} &\\
& \times {e^{-\rho \kappa_s \tau ((s+x)^{1-\alpha} \mathsf{E}_G[\nu] - s^{1-\alpha}\mathsf{M})/(1-\alpha)}}\} dx&\nonumber\\
& + \tfrac{1}{{\lambda _s^0 + \rho \kappa_s^0 \tau \mathsf{E}_G[\nu]}} {e^{ - \left( {{{{T_t}}^{1 - \alpha }} - {{s}^{1 - \alpha }}} \right){\lambda _s}/(1 - \alpha ) }}&\nonumber\\
& \times e^{-\rho \tau \kappa_s (T_t^{1-\alpha}\mathsf{E}_G[\nu]-s^{1-\alpha}\mathsf{M})/(1-\alpha)} ]&\nonumber\\
& + {1_{s > {T_t}}}[\tfrac{e^{-s \rho \tau \kappa_s^0 (\mathsf{E}_G[\nu]-\mathsf{M})}}{{\lambda _s^0 + \rho \kappa_s^0 \tau \mathsf{E}_G[\nu]}}]&\nonumber
\end{flalign}
which only requires a Riemann sum over $[0,T_t-s]$ for $s<T_t$. Considering the small length of this region and common \textit{base failure rates} for electrical components, a sum with step length of $10$ to $100$ hours can provide a very accurate result for the MRL. $\mathsf{M}$ can be either neglected (assuming that the mean of all fluctuations in the life time of the component is zero) or for accurate results, can be derived by calculating the mean error of the stress factor over all times (i.e. $\mathsf{M}=\sum \nu(k) / N$ where $\nu(k)=v(k)-v^*$ is the observed voltage fluctuations at the time $k$ and $N$ is the total number of observations). On the other hand, $\mathsf{E}_G[\nu]$ can be calculated as a moving average over these fluctuations. The moving average window is a design parameter. One can consider a day or a week for the window of this moving average. This window claims that the expected mean of the future events based on the newly observed stress factors should be similar to what we have observed in the past day or week. One can select $\mathsf{E}_G[\nu]=\mathsf{M}$ assuming that the mean of all events in the future is the mean of all past events. (\ref{_eqacl}) can be updated in the same manner. One can use the new voltage and temperatures to update this parameter. On the other hand, to observe a smooth model, one can use the average of voltages and temperatures over time. Therefore, at a time $t$
\begin{flalign}
\label{_calcM}
&\mathsf{M}={(\#\{\nu_0,...,\nu_t\})}^{-1} \sum\limits_{k=0}^{t} {\nu(k)}&\\ 
&\mathsf{E}_G[\nu]=N^{-1} \sum\limits_{k=t-N-1}^{t} {\nu(k)}&\\ 
\label{_calcA}&{{\bar A}_a}(n) = 0.47{e^{\tfrac{{3 \sum {N^{-1}\nu(k)}}}{{v_{CE}^*}} + \tfrac{{0.4eV}}{{{K_b}}}(\tfrac{1}{{313}} - \tfrac{1}{{\sum {N^{-1}T_a(k)}}})}}
\end{flalign}
where $\#$ is the cardinality of the set of all samples.
\section{Case Study}
\begin{figure}[tb]
\centering
\includegraphics[trim = 0in 0in 0in 0in, clip, width=2.7in]{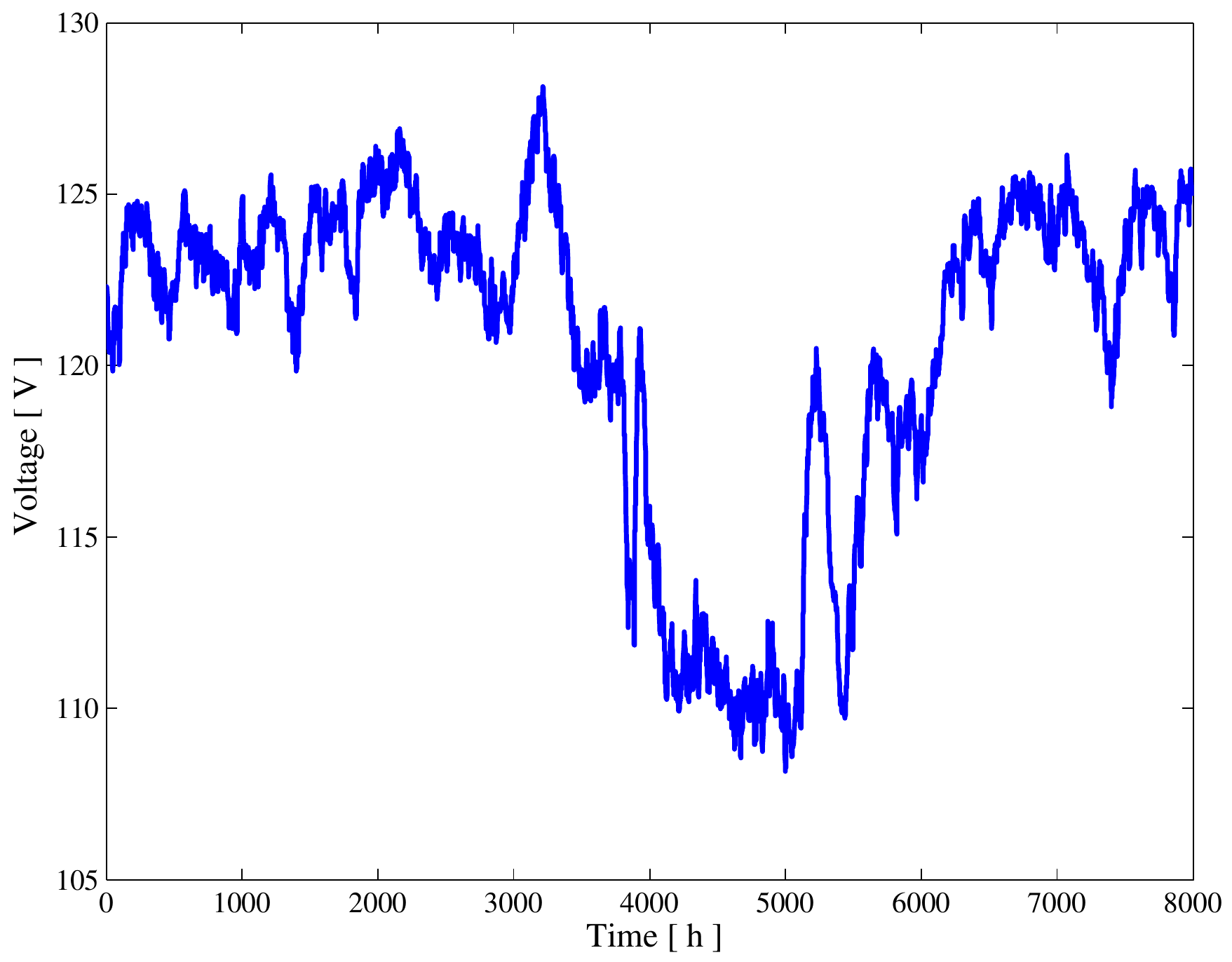}
\caption{The measured voltage of the battery.} 
\label{_ex_vb}
\end{figure}

\noindent In this section, an experimental case study is performed using the proposed on-line survival analysis method. In this example, the survival of the input capacitor of a dc-dc converter under variations of the voltage of the source is studied. Consider a dc-dc converter that is supplied from a 120 V battery pack. The input capacitor of this converter is directly connected to the battery pack and will observe the voltage variations of the pack. In order to expedite the measurements, the data is collected every 1 minutes. However, the analysis is performed assuming that the data is collected at each hour. This expedites the measurements from 1.5 years to only 6 days. The measured voltage of the battery is shown in Fig. \ref{_ex_vb}. It can be observed that based on the state of charge of the battery, the voltage of the battery varies between 110 V to 130 V. 
\begin{figure}[tb]
\centering
\includegraphics[trim = 0in 0in 0in 0in, clip, width=2.7in]{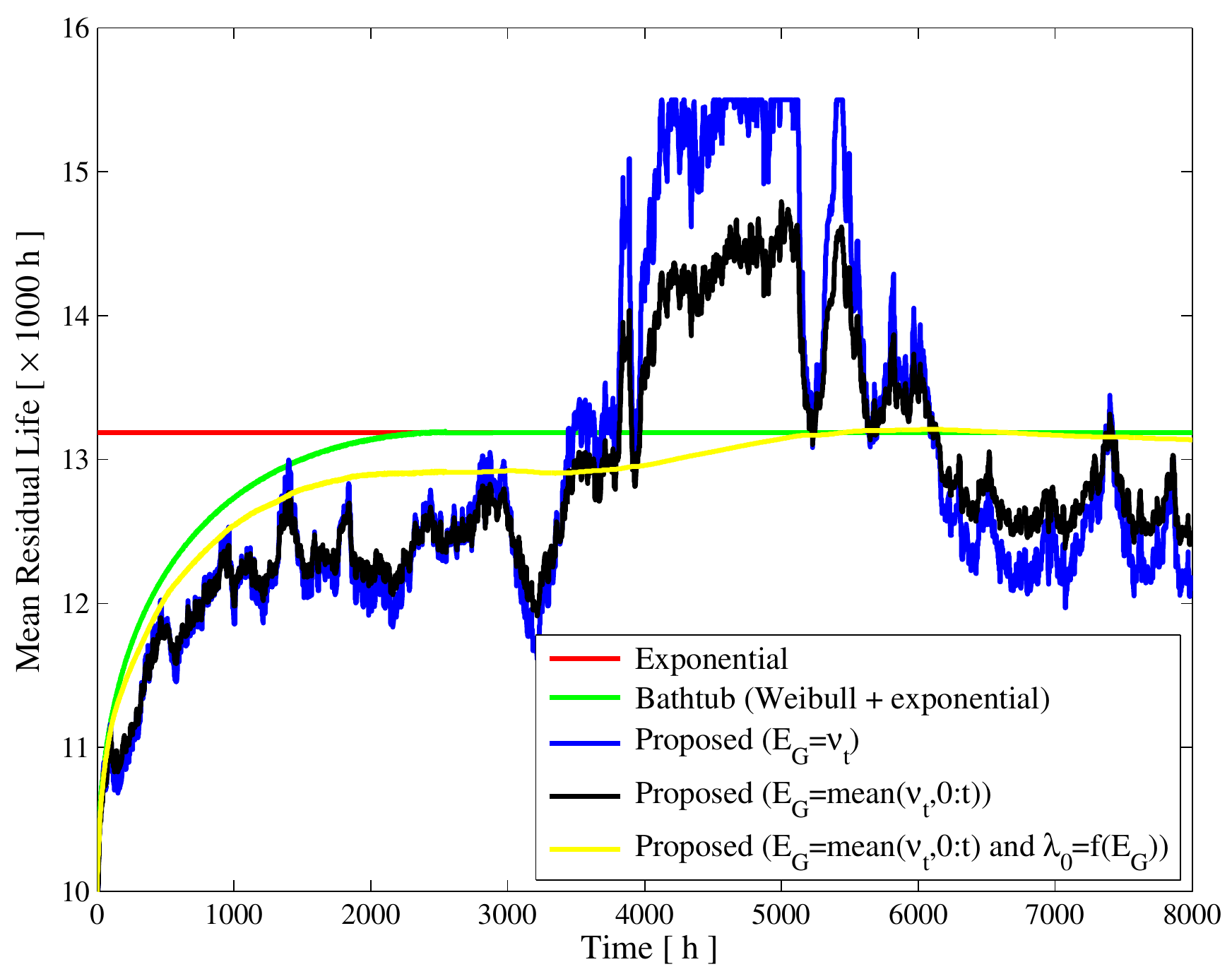}
\caption{Mean residual life of the capacitor.} 
\label{_ex_mrl}
\end{figure}

In the conventional methods for estimation of the survival of the capacitor, the nominal battery voltage is used to calculate the distribution functions as well as MRL. However, in this paper, the function is calculated based on the measured voltage of the battery at each measurement cycle. Consider a case that the battery has been depleted for a long time. Conventional methods will consider the capacitor to be aging with a constant acceleration factor that corresponds to 120 V. However, the proposed on-line estimation method will calculate the failure rate based on the measured voltage of the battery. Hence, the model will use a smaller aging factor for this case.

The on-line MRL estimation for the capacitor based on the measured voltage of Fig. \ref{_ex_vb} is shown in Fig. \ref{_ex_mrl}. This figure illustrates the difference between on-line and off-line MRL estimation methods. The off-line methods including a constant failure rate model and the bathtub model do not consider the variations of the voltage of the battery. However, the on-line method will use the data collected up to the measurement time to calculate the failure rate function in the model. Moreover, the expected failure rate function (of the future) is estimated using the Cox process. Therefore, the actual age of the capacitor is estimated more accurately. In order to study various applications of the proposed model, 3 candidate models are illustrated. In all of these models, $\mathsf{M}$ is calculated as (\ref{_calcM}). In the first model (illustrated in blue), the expected mean of voltage stresses in the future (i.e. $\mathsf{E}_G[\nu]$) is simply set to the last observed voltage. This approach is simple and can provide a good estimation for some applications. However, a more accurate approach is to calculate this mean as a mean of all observed voltages. This approach is illustrated in black. In the last model (yellow), in addition to a mean for calculation of $\mathsf{E}_G$, this mean is also used to calculate the base aging factor as in (\ref{_calcA}). It should be noted that the MRL has a maximum limit which is based on mechanical/chemical wear-out. Hence, even if no voltage is applied to this capacitor, the MRL will no go higher than 15500hours. This is the reason for the clamping effect in model 1 in this figure.

\begin{figure}[tb]
\centering
\includegraphics[trim = 0in 0in 0in 0in, clip, width=2.7in]{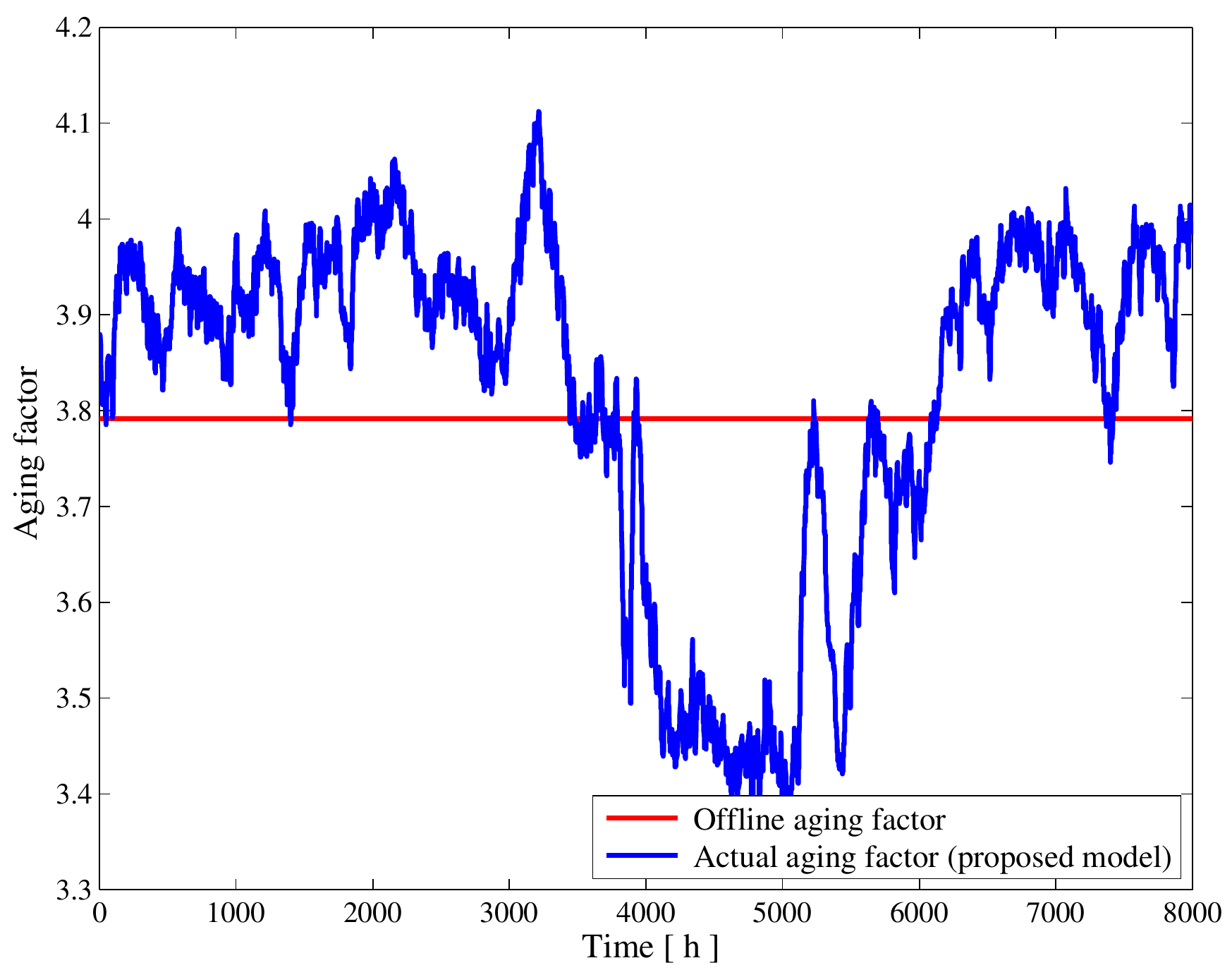}
\caption{On-line calculation of the capacitor aging factor.} 
\label{_ex_ag}
\end{figure}

The aging factor (acceleration factor) for the capacitor is shown in Fig. \ref{_ex_ag}. It can be observed that the average aging factor is 3.7. This number is used in conventional methods. This number shows that the capacitor is 3.7 times older than the calendar time. In this paper, aging factor is updated at each sample. For instance, Fig. \ref{_ex_ag} illustrates aging factor for models 1 and 2 in the above discussion. Therefore, the infant mortality region will end in 2500 h as it can be observed from Fig. \ref{_ex_mrl}.
\begin{figure}[tb]
\centering
\includegraphics[trim = 0in 0in 0in 0in, clip, width=2.7in]{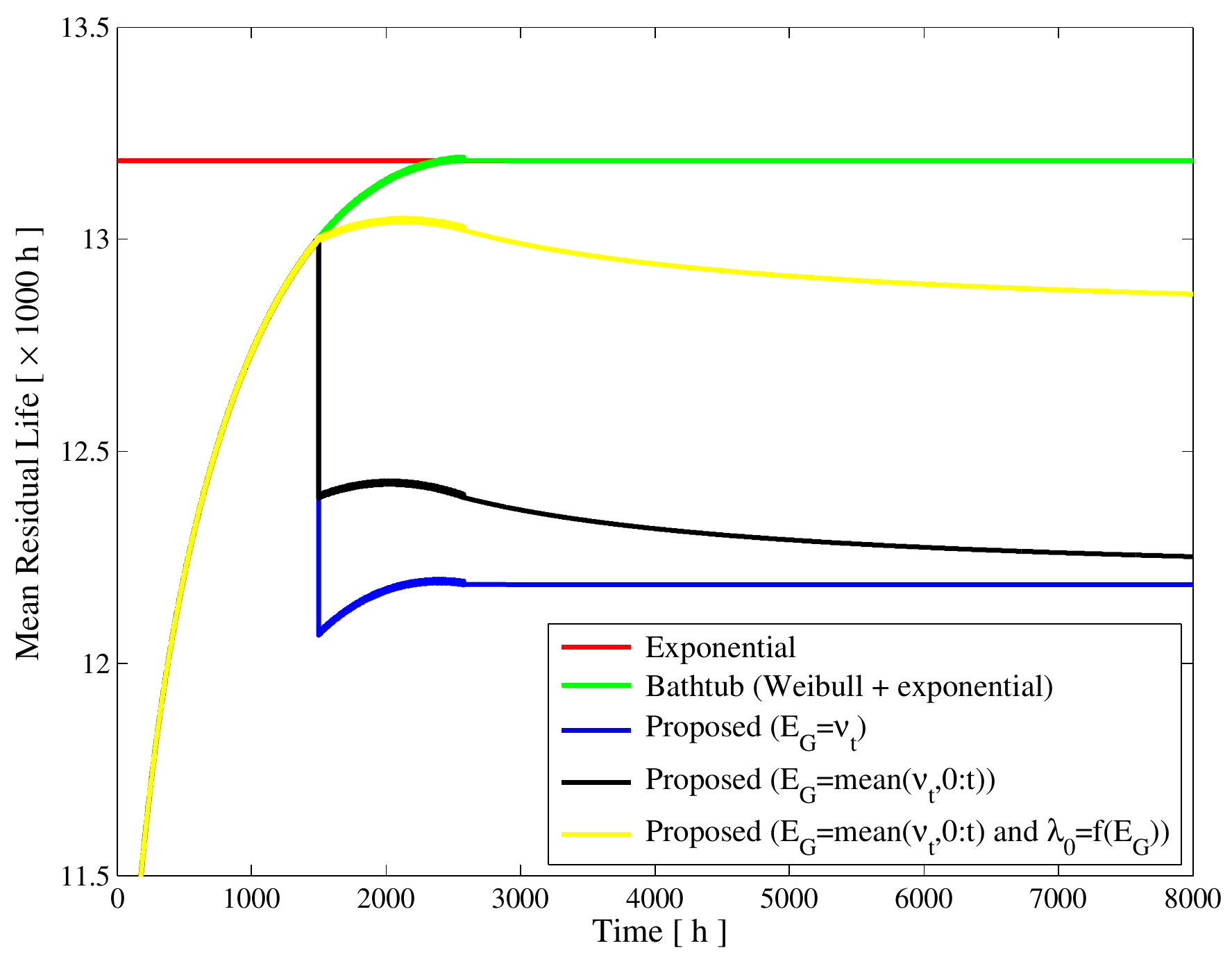}
\caption{Step change comparison for various models.} 
\label{_exC}
\end{figure}

A direct comparison between exponential, Weibull, and the proposed three example models, is shown in Fig. \ref{_exC}. In this example, the voltage of the battery is 120V till 1500h. Then, this voltage is increased to 125V and kept constant. Response of each model can be observed from this figure. Model 1 has the fastest response while model 3 has the slowest response. At $t\rightarrow \infty$, the proposed models converge.
\section{Applications and Further Research}
\noindent Although this model was studied for survival analysis of electrical components, the same methodology can be incorporated to derive a distribution function for availability of energy from renewable energy sources. 

On the other hand, the moving average used for calculation of the parameters in this paper can be replaced with a more accurate estimator to get a better sense of the underlying distribution functions.
\section{Conclusions}
A method for on-line calculation of the mean residual life of electrical components was introduced in this paper. The proposed method is based on applications of Cox processes in modeling the probability of the failure of components under stochastic variation of operational conditions. This method is applicable to various reliability models. In this paper, this method was developed for reliability analysis using a bathtub failure rate function. Using this method, expected residual life of a component was derived. Due to on-line observation of the stress levels, this method provides more accurate results in estimating the age of the system and predicting the probability of failures. Moreover, by tuning the stochastic failure rate function based on the observed stress levels, Cox processes can provide a good models for survival analysis of various components. This paper presented an experimental case study to demonstrate the application of the proposed method.

\bibliographystyle{IEEEtran}
\bibliography{dc}
\end{document}